\documentclass[submission,copyright,creativecommons]{eptcs}
\providecommand{\event}{FMAS 2022} 
\usepackage{underscore}           
\usepackage{cite}
\usepackage{amsmath,amssymb,amsfonts}
\usepackage{graphicx}
\usepackage{textcomp}
\usepackage{xcolor}
\usepackage{circle}
\usepackage{listings}
\usepackage{paralist}
\usepackage{subfig}
\def\BibTeX{{\rm B\kern-.05em{\sc i\kern-.025em b}\kern-.08em
    T\kern-.1667em\lower.7ex\hbox{E}\kern-.125emX}}
    
\usepackage[textsize=normalsize,textwidth=3cm]{todonotes}
\usepackage{macros}

\usepackage{hyperref}

\title{Extending Attack-Fault Trees with Runtime Verification\thanks{Part of Cardoso's and Fisher's work was supported by The Royal Academy of Engineering under the Chair in Emerging Technologies scheme and by EPSRC under project EP/V026801 (Trustworthy Autonomous Systems ``Verifiability Node'').}}

\author{Rafael C. Cardoso
\institute{Department of Computing Science\\ University of Aberdeen\\ Aberdeen, United Kingdom}
\email{rafael.cardoso@abdn.ac.uk}
\and
Angelo Ferrando
\institute{DIBRIS\\ University of Genova\\ Genova, Italy}
\email{angelo.ferrando@dibris.unige.it}
\and 
Michael Fisher
\institute{Department of Computer Science\\ The University of Manchester\\ Manchester, United Kingdom}
\email{michael.fisher@manchester.ac.uk}
}

\def\titlerunning{Extending Attack-Fault Trees with Runtime Verification}
\def\authorrunning{R.C. Cardoso, A. Ferrando \& M. Fisher}

\begin{document}

\maketitle

\begin{abstract}
Autonomous systems are often complex and prone to software failures and cyber-attacks.
We introduce RVAFTs, an extension of Attack-Fault Trees (AFTs) with
runtime events that can be used to construct runtime monitors. These
monitors are able to detect when failures, that can be caused either by an attack
or by a fault, occur. The safety and security properties monitored
are, in turn, derived from the hierarchical decomposition of RVAFTs.
Our approach not only provides further use of AFTs, but also improves
the process of instrumentation often required in runtime verification.
We explain the principles and provide a simple case study demonstrating
how RVAFTs can be used in practice. Through this we are also able to
evaluate the detection of faults and attacks as well as assessing the
computational overhead of the monitors.

\end{abstract}

\section{Introduction}

Safety and security are both integral parts of system development. Usually these aspects are approached separately and, consequently, have vast literature addressing solutions for them individually. In recent years, communities from both research areas have been converging to develop approaches that combine safety and security. Considering both of these concerns together is most useful for critical applications such as robotic and autonomous systems~\cite{robotics10020067}. The robot or vehicle may have prescribed actions to take to recover from a specific situation (stop, return to base, reset), but needs to recognise attacks and failures in order to invoke the appropriate action.

Attack-Fault Trees (AFTs)~\cite{aft} are an example of a formalism that combines attack trees (security) and fault trees (safety) into a single representation. The hierarchical decomposition of the tree produces paths that eventually lead to attack or fault nodes. This combined representation of security threats and safety issues makes AFTs useful in guiding the design and development of robotic systems, where both concerns are often of paramount importance. 

Runtime Verification (RV) is a formal verification technique~\cite{DBLP:journals/jlp/LeuckerS09} for checking the behaviour of the system under analysis at execution time. To achieve this, monitors are generated from formally specified properties (the requirements to be verified). A runtime monitor constantly analyses traces of events that are generated during the system's execution to assess the satisfaction or violation of a property. Because RV is focused on analysing the system at runtime, it is non-exhaustive (not all possible states are explored) since the monitor can only give a verdict based on the execution traces that have been observed. 


Using a representation that considers both security and faults can reduce the steps required for successful verification and validation of the system. Specifically to RV, generating monitors for safety and security concerns from the same representation allows better upkeep of the monitors, provides better traceability of violations, improves the instrumentation process of runtime monitors, and provides a route to closer interactions between safety and security in the future.

In this paper, we focus on identifying attacks and faults at runtime and how runtime monitors can detect them using different branches of an AFT. We do not cover the steps after detection, which would involve mitigating, correcting, or recovering from attacks or faults.\footnote{We note that the IEEE P7009 standard being developed will specifically address \emph{fail-safe} aspects of autonomous systems~\cite{P7009:21}.} Our contributions comprise: an extension of AFT with runtime events (called RVAFT), a translation between RVAFTs and a property specification language for runtime monitors, and a demonstration of how these runtime monitors can be generated in a case study. 


\section{Related Work}
\label{sec:rw}


A RV approach for detecting fault injection attacks is presented in~\cite{Kassem19}. Even though the authors only consider security activities, their work is relevant because the literature on RV for detecting attacks is quite limited when compared to RV for safety. Specifically, they define formal models of runtime monitors that can detect fault injections resulting in test inversion and arbitrary jumps attacks. The monitors are expressed as Quantified Event Automata.



The \emph{R2U2} framework~\cite{DBLP:journals/fmsd/MoosbruggerRS17} is an approach that uses RV for monitoring security properties and diagnosing security threats. The framework provides a comprehensive suite of techniques for both software and hardware real-time monitoring targeting unmanned aerial systems. Some case studies are shown on the NASA DragonEye system as a proof-of-concept As correctly reported in their work, it is not possible to identify an attack at execution time simply by monitoring the runtime events, since the cause could be related to a myriad of other situations (such as faulty hardware or software). It is necessary to carry out in depth post-attack analysis to match the behaviour of the system during and after the suspicion of an attack. Their work however is tailored specifically to unmanned aerial systems and is not generally applicable to other types of systems.




A recent systematic literature review~\cite{survey1} investigated approaches that provide co-analysis of safety and security in the early stages of system development. In their review, the relevant publications are categorised according to how safety and security were related to each other: (a) they can be combined, where the mutual influence between safety and security is considered; (b) safety-informed security, where the influence of safety on security is considered; and (c) security-informed safety, where the influence of security on safety is considered. Most papers were categorised as (a), with (b) as a close second, and none in (c). The paper about AFTs~\cite{aft} was placed into the (a) category.

AFTs were first introduced in~\cite{aft} to combine safety and security concerns together into a single representation capable of identifying events that lead to attacks and faults. As such, it is essentially based on previous research that was carried out separately into fault tree analysis and attack tree analysis. The work in~\cite{aft} not only introduced AFTs, but also provided a technique to analyse them using statistical model checking by annotating the tree's nodes with the relevant information required to perform the analysis. In our work, we do not make use of these annotations, instead we extend AFTs to be used for generating runtime monitors. However, this does not preclude the possibility of performing such analysis, as long as the annotations are added, statistical model checking can still be performed following the techniques described in~\cite{aft}. 

An AFT is a directed acyclic graph, where the top-level node (i.e., the root) is the undesired event that can be triggered by decomposing the tree. In other words, a branch follows a string of causalities that eventually terminates in one or more leaves. The nodes in the tree are connected by the use of logical gates. These gates control what branch is being considered depending on the current situation of the child nodes that are directly connected to the gate. In~\cite{aft}, seven different types of gates are described, coming from existing attack or fault trees literature. 
In our work, we only consider four of these logical gates that were the most suited for RV.
The first two gates are the traditional $\mathit{AND}$ and $\mathit{OR}$ gates for conjunction and disjunction (respectively). Note that the basic $\mathit{AND}$ gate does not impose any ordering restriction between the nodes. To express ordered conjunction we can use the $\mathit{SAND}$ (Sequential AND) gate, with two variants: left to right, or right to left. Lastly, we have the $\mathit{VOT}$ (Voting OR) gate for when we need to say that at least $k$ nodes out of $n$ have to occur to trigger the gate. The remaining gates ($\mathit{PAND}$, $\mathit{FDEP}$, and $\mathit{SPARE}$) were not relevant for the generation of runtime monitors in our case study, but may be further investigated in future work.


AFTs are generated at early stages in the development of a system to pinpoint possible security and safety issues that the system may have. After they are created, they remain useful throughout all development stages and can be used to inform which critical areas of the system require more work in order to avoid or mitigate potential security threats and safety concerns. Another important aspect to note about AFTs is that they do not explicitly relate to time. New gates could be introduced to add temporal capabilities, such as a simultaneous $\mathit{AND}$ gate (nodes have to happen at the same time) and/or gates with time intervals, but we leave these extensions for future work. For the purposes of runtime monitoring in this paper, we assume that these events can be eventually observed at some point in time.

Before AFTs there have been a number of approaches that have been attempted to combine security and safety concerns at early stages of software development ---  Fault Tree Analysis (FTA)~\cite{BROOKE2003256}, Extended Fault Tree (EFT)~\cite{NAIFOVINO20091394}, and Failure-Attack-CounTermeasure (FACT)~\cite{Sabaliauskaite15} were all precursors to AFTs.

\section{Extending Attack-Fault Trees with Runtime Events}
\label{sec:ext}

In this section, we introduce RVAFT, an extension of AFT that supports the description of runtime events for creating monitors. The addition of these events allows the generation of runtime monitors based solely on the RVAFT. The security and safety properties can then be formulated by hierarchical decomposition of the RVAFT. This process is guided by the logical gates connecting the nodes in the tree. Note that while we focus on AFTs for this paper, the methodology presented here could be applied to any tree representation of attacks and/or faults as long as the standard logical gates are semantically equivalent.

Creating runtime monitors using RVAFTs has two major advantages: (a) runtime monitors can be distinctively created to verify, at runtime, safety and security properties based on a single representation; and (b) the process of relating observable runtime events to the property we want to verify is seamless. Point (a) may seem less relevant if we assume that we already have separate properties for both safety and security with the relevant observable events that can identify faults and attacks (respectively). However, in practice, this is rarely the case, since without the proper representation (such as an AFT) it can be very difficult to differentiate between faults and attacks at design time. Point (b) is related to the difficulties in associating observable runtime events to a specific property, which does not occur when using RVAFTs since properties and events are intrinsically connected.

\begin{figure}[ht]
\centering
\framebox{
\includegraphics[width=0.35\linewidth]{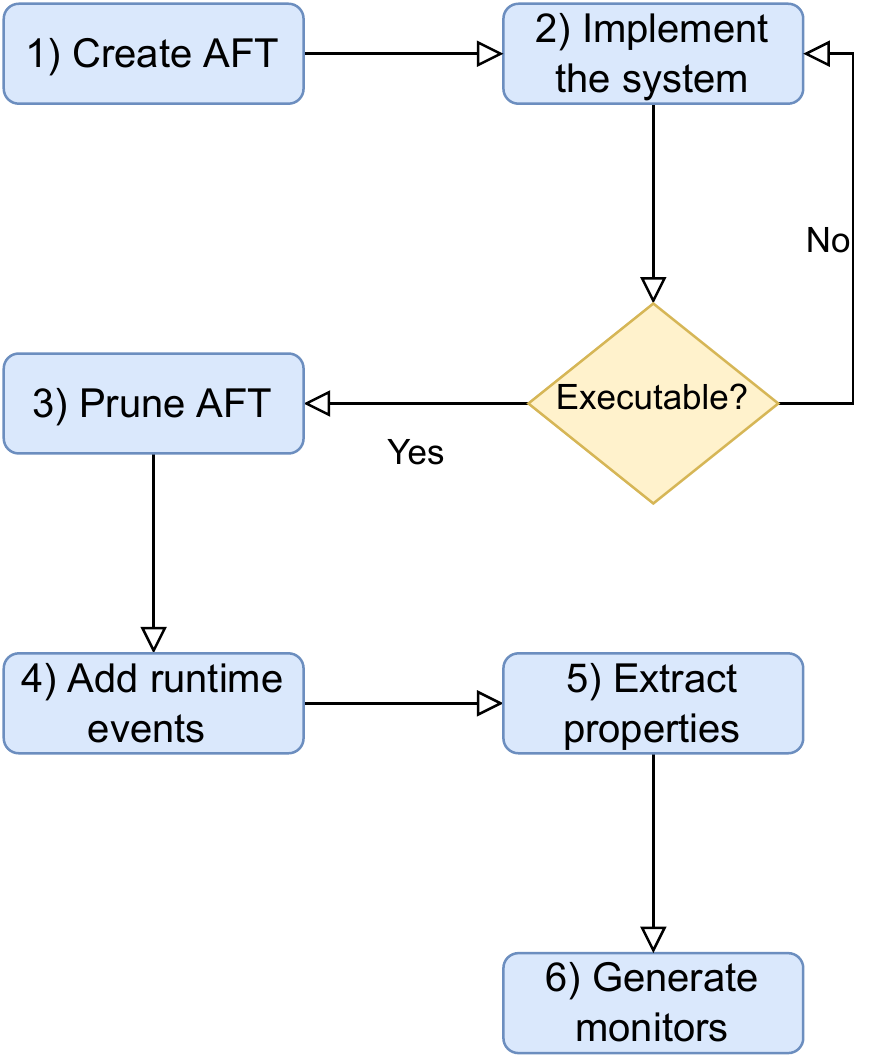}
}
\caption{Flowchart for converting AFTs to RVAFTs and generating runtime monitors out of them.}
\label{fig:flowchart}
\end{figure}

To apply our approach we use the development process illustrated in Figure~\ref{fig:flowchart} and detailed as follows:\footnote{We omit software development steps that are not directly related to performing runtime verification using AFTs.}
\begin{enumerate}
    \item \label{step1} Create an AFT for the system.
    \item \label{step2} Continue to develop the system until either a prototype, simulation, or application can be executed.
    \item \label{step3} Prune the AFT until it has only security and safety concerns that the developer wants to detect at runtime.
    \item \label{step4} Extend the AFT with runtime events for each node (excluding the root) based on either runtime logs of the system or expert knowledge from the system's developers.
    \item \label{step5} Extract safety and security properties from the decomposition of the RVAFT.
    \item \label{step6} Generate runtime monitors based on the RVAFT and the extracted properties.
\end{enumerate}
\noindent Step~(\ref{step1}) is based on the work presented in~\cite{aft} and step~(\ref{step2}) is outside the scope of this paper. Thus, we will focus on discussing steps (\ref{step3})--(\ref{step6}).


Step~(\ref{step3}) is optional and can be skipped if there is nothing to prune, since it may be the case that we want to generate monitors based on the entire AFT. However, AFTs tend to be very general and aim to represent many of the security and safety concerns that seem relevant during the design phase, but may lose importance (or even become obsolete) when analysing the implemented system. It is also possible that the existence of some attacks and faults have already been verified using models of the system (e.g., model checking, simulation-based testing, etc.) and no longer require further verification. Lastly, it may be required that we prioritise which concerns really \emph{need} to be verified at runtime, since monitors may add an overhead (the severity of the overhead depends on the hardware and system specifications, as well as the RV technique being used) which can lead to performance issues. 
%


In Step~(\ref{step4}), each node in the pruned AFT (except for the root) has to be assigned a runtime event that represents the event wherein the node can be observed. This should be straightforward for most nodes and can be done by checking execution logs of the system or may alternatively be based on expert advice from the system's engineers. Some nodes may need to be refactored depending on how the event can be observed at runtime. 

In Step~(\ref{step5}) we should have a complete RVAFT. At this step, we simply perform hierarchical decomposition of the tree to generate safety and security properties that already include, by default due to the previous step, the observable events related to these properties. From this, a monitor can be easily generated to verify these properties at runtime in Step~(\ref{step6}). Steps~(\ref{step3}) and~(\ref{step4}) are manual, Step~(\ref{step5}) is semi-automated (automated translations for the gates are provided, but because there is no formal language to represent these trees the extraction has to be done manually), and Step~(\ref{step6}) is fully automated.

To exemplify the use of our approach we introduce a remote inspection case study in Section~\ref{sec:case-study} that we use to demonstrate how to perform Steps~(\ref{step1}), (\ref{step3}), and (\ref{step4}) in Section~\ref{sec:case-aft}, Step~(\ref{step5}) in Section~\ref{sec:properties}, and Step~(\ref{step6}) in Section~\ref{sec:monitor}. Before we discuss the case study, we present a brief overview of the property specification language that we have chosen to use as an example and how to represent the AFTs logical gates in this language.

\subsection{RML}

Runtime Monitoring Language\footnote{\url{https://rmlatdibris.github.io/} (Accessed: 08/09/2022)} (RML) is a Domain-Specific Language (DSL) for specifying non context-free properties to be used in RV. We use RML in this paper for its support of parametric specifications and its implementation that is available for robotic applications. We provide a simplified and abstracted view of RML, with the complete syntax and semantics of RML available in~\cite{AnconaFFM21}.


An RML property is a tuple $\langle t,ETs \rangle$, with $t$ a term and $ETs=\{ET_1,\ldots,ET_n\}$ a set of event types. An event type $ET$ is represented as a set of pairs $\{k_1:v_1,\ldots,k_n:v_n\}$, where each pair identifies a specific piece of information ($k_i$) and its value ($v_i$). An event $Ev$ is denoted as a set of pairs $\{k_1':v_1',\ldots,k_m':v_m'\}$. Given an event type $ET$, an event $Ev$ matches $ET$ if $ET \subseteq Ev$, which means $\forall (k_i:v_i) \in ET \cdot \exists (k_j:v_j) \in Ev \cdot k_i = k_j \land v_i = v_j$. 

An RML term $t$, with $t_1$, $t_2$ and $t'$ as other RML terms, can be (note that we only describe terms that were used in our case study):
\begin{compactitem}
    \item $ET$, denoting a singleton set containing the events $Ev$ such that $ET \subseteq Ev$;
    \item $t_1 \;\; t_2$, denoting the sequential composition of two sets of terms;
    \item $t_1 \; | \; t_2$, denoting the unordered composition of two sets of terms;
    \item $t_1 \lor t_2$, denoting the union of two sets of terms;
    \item $\{ let \; x;\ t' \}$, denoting the set of terms $t'$ where the variable $x$ can be used.
\end{compactitem}



\subsection{Logical Gates RML Translation}

\begin{figure}[!htbp]
\centering
\scalebox{0.8}{

\subfloat[$\mathit{OR}$ gate.]{\includegraphics[width=0.3\textwidth]{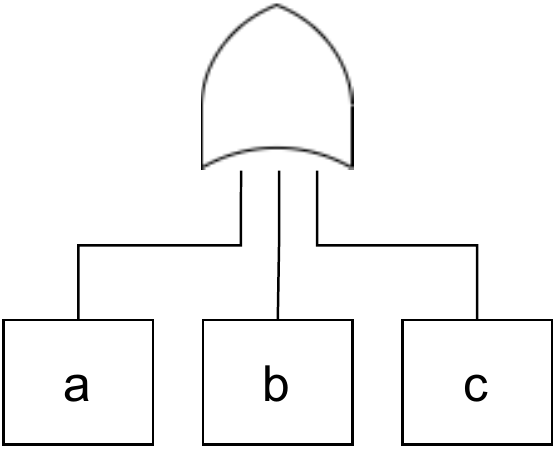}\label{fig:rvaftor}}\hfill\qquad
\subfloat[$\mathit{VOT}$ gate.]{\includegraphics[width=0.3\textwidth]{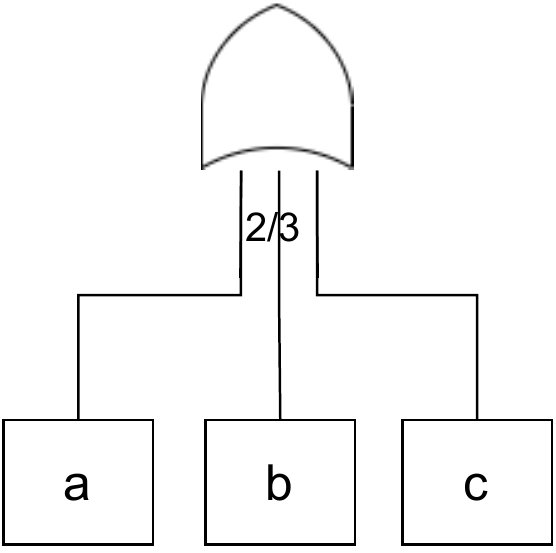}\label{fig:rvaftvot}}\hfill\qquad
\subfloat[$\mathit{AND}$ gate.]{\includegraphics[width=0.3\textwidth]{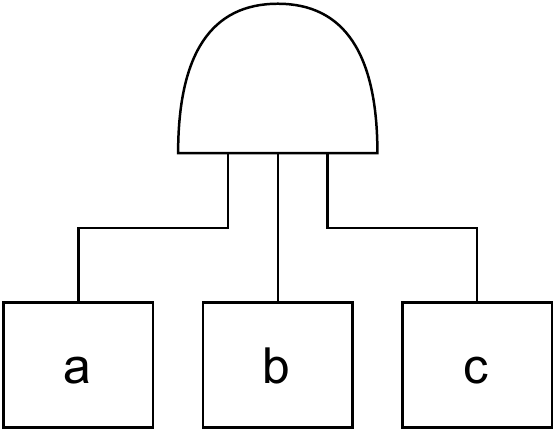}\label{fig:rvaftand}}%
}

\scalebox{0.7}{
\subfloat[$\mathit{SAND}$ gate left to right.]{\includegraphics[width=0.32\textwidth]{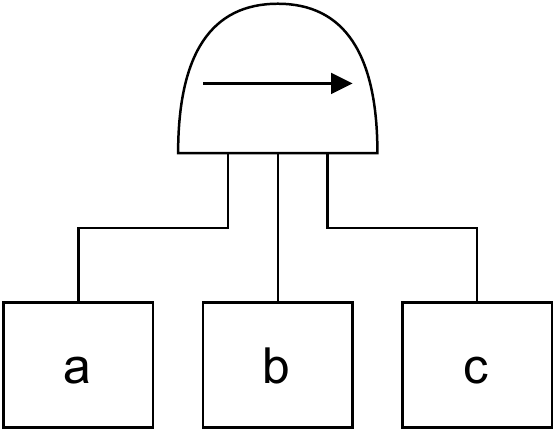}\label{fig:rvaftsand1}}%
\qquad%
\subfloat[$\mathit{SAND}$ gate right to left.]{\includegraphics[width=0.32\textwidth]{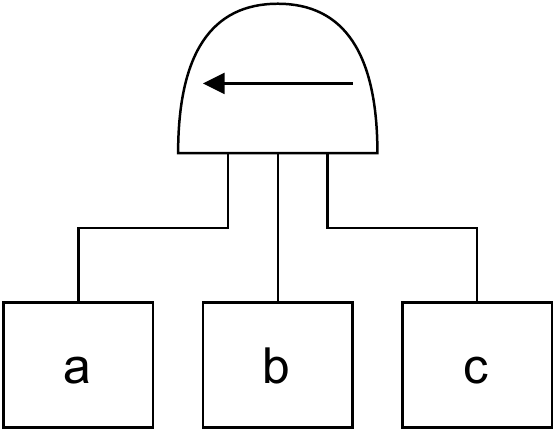}\label{fig:rvaftsand2}}
}
\caption{Abstract gates.}\label{fig:gates}

\end{figure}

The four logical gates that we translate to RML are shown in Figure~\ref{fig:gates} with each gate containing three abstract nodes.

\subsubsection*{OR Gate Translation}

Starting from the abstract example in Figure~\ref{fig:rvaftor}, we can translate the disjunction $\mathit{OR}$ gate to obtain the following RML term ($\tau$):
$$
\begin{array}{l}
\tau = a \;\lor\; b \;\lor\; c
\end{array}
$$
which is simply the union of all possible events.


\subsubsection*{VOT Gate Translation}

In the $\mathit{VOT} (k/n)$ gate ``at least $k$'' over $n$ events have to be observed. This can be achieved in RML by denoting a hierarchy of terms, where in each term a different event is processed individually. For example, if we consider the example in Figure~\ref{fig:rvaftvot} with events $a$, $b$ and $c$, with $k = 2$ and $n = 3$, then the corresponding RML specification would be as follows:  
$$
\begin{array}{l}
\tau = (a \;\tau_{bc}) \;\lor\; (b \;\tau_{ac}) \;\lor\; (c \;\tau_{ab})\\
\tau_{bc} = (b \;\lor\; c) \\
\tau_{ac} = (a \;\lor\; c) \\
\tau_{ab} = (a \;\lor\; b)
\end{array}
$$
each line describes a different RML term. In $\tau$ we have the main term, with three options available: the events $a$, $b$, or $c$. Depending on which of these events is actually observed, the terms that follow are determined. For instance, if in $\tau$ the observed event is $a$, then the following term becomes $\tau_{bc}$, where the $b$ or $c$ events are required to be observed. This is symmetric for the other events, where a different event is observed in $\tau$, and one of the other two is observed later on.


This flat behaviour, where the counting of the observed events is obtained by explicitly unrolling the RML terms is necessary to keep track of which events have been observed in the past, and which ones still need to be observed in the future. RML offers mechanisms to keep a  memory of past events, but in its current implementation it does not allow the use of set theory. If the RML event types definition supported sets, then a more elegant and compact representation of the flat terms above could be obtained. If that were possible, then the term could be represented as a parametric term with a set of past events as input to be checked for discarding events that were already seen.

Note that most logics and DSLs (including RML) could represent this simple example as: $ab \,\lor\, ac \,\lor\, ba \,\lor\, bc \,\lor\, ca \,\lor\, cb$. However, as $k$ and $n$ grow it will become increasingly difficult to represent this succinctly. Using RML terms, this representation becomes much more human-readable and easier to expand, while also remaining independent on the values of $k$ and $n$.
 
\subsubsection*{AND Gate Translation}

The $\mathit{AND}$ gate may initially seem as simple as the disjunction gate, but it has one important detail that makes it different from traditional conjunction representations in most standard logics.  In AFTs, $\mathit{AND}$ is unordered, meaning that in the abstract example from Figure~\ref{fig:rvaftand} we can observe the events $a$, $b$ and $c$ in any order.

Thus, a flat translation of this gate in traditional logics (such as Linear-time Temporal Logic (LTL)~\cite{DBLP:conf/focs/Pnueli77}) would be: $(a \, \land  \, b  \, \land  \, c)  \, \lor  \, (a  \, \land  \, c  \, \land  \, b)  \, \ldots  \, (c  \, \land  \, b  \, \land  \, a)$. That is, all possible permutations of the observable events related to the gate. However, as we have seen, RML supports the parallel composition operator ($\,|\,$), which allows us to specify the unordered intersection of traces. Using this operator we can derive the following RML term:
$$
\begin{array}{l}
\tau = a \;|\; b \;|\; c
\end{array}
$$
which means that we can observe any possible ordering between the events.

\subsubsection*{SAND Gate Translation}

The Sequential $\mathit{AND}$ ($\mathit{SAND}$) is a sequential composition, denoting that events have to be observed in a particular order. This gate represents the traditional conjunction often found in logics. Figure~\ref{fig:rvaftsand1} shows the case where events have to be observed from left to right, while the opposite ordering (right to left) is shown in Figure~\ref{fig:rvaftsand2}. Since these are concatenations of events, the translation in RML is straightforward:
$$
\begin{array}{l}
\tau = a \;\; b \;\; c\\
\tau' = c \;\; b \;\; a
\end{array}
$$
with $\tau$ being the RML term for $\mathit{SAND}$ left to right and $\tau'$ the term for $\mathit{SAND}$ right to left (remember that the space in RML denotes concatenation).


\section{A Remote Inspection Case Study}
\label{sec:case-study}

Our case study is based on a 3D simulation of a Jackal\footnote{\url{https://clearpathrobotics.com/jackal-small-unmanned-ground-vehicle} (Accessed: 08/09/2022)}, a four-wheeled unmanned ground vehicle (referred to as the `rover' from now on), coupled with a simulated radiation sensor that is used to take radiation readings of points of interest in a nuclear facility. This simulation is based on the work presented in~\cite{robotics10030086}, which explains how the simulated sensor works and how radiation was simulated in the environment. In our version of the simulation the rover is autonomously controlled by a rational/intelligent agent~\cite{WooldridgeRao99:book}. The simulation is developed in the Robot Operating System\footnote{\url{https://www.ros.org/} (Accessed: 08/09/2022)} (ROS), the \textit{de facto} standard middleware for the development of robotic applications. ROS is based on modular components called `nodes' that can communicate with each other through message passing using a publish/subscribe architecture for a specific communication channel (also called a \emph{topic}). Runtime monitors for ROS applications often rely on observing these messages in order to verify a property at runtime.

A typical mission in our simulation starts with the rover positioned at the entrance of a nuclear facility. The goal of this mission is to inspect a number of points of interest (i.e., waypoints). Inspecting a waypoint serves two purposes: taking radiation readings to check if the radiation is at an acceptable level, and using a camera to detect abnormalities such as leakage in barrels and pipes. After inspecting all of the waypoints, the rover can either return to the entrance to await for a new mission, or keep patrolling and inspecting the waypoints. If radiation is detected to be too high ($\geq 250$, an arbitrary number of radiation units that was selected for demonstration purposes), then the rover should immediately stop whatever it is doing and move to the entrance of the facility to undergo decontamination procedures. 

For the purposes of this paper, the implementation details of the simulation and the autonomous agent are not relevant. Our runtime monitors do not require a model of the system, they only need to be able to observe events that are generated at runtime, i.e., we can consider the implementation as a black-box. 


\subsection{Case Study AFT}
\label{sec:case-aft}

\begin{figure}[htbp]
	\centering
	\subfloat[This AFT concerns the undesirable event of when a robot can be damaged by radiation.]{\includegraphics[width=0.54\textwidth]{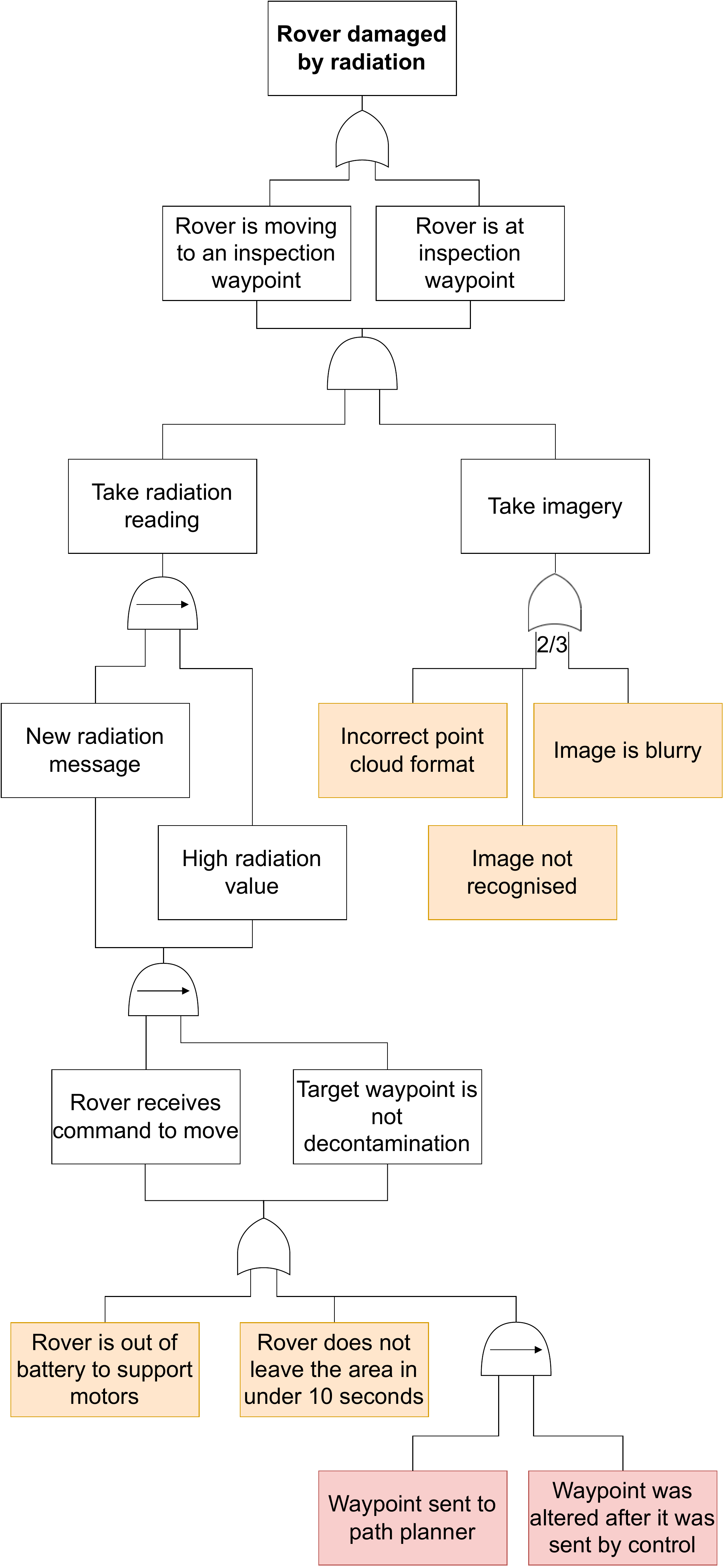}\label{fig:aft}}%
    \hspace{0.15cm}\vrule{}\hspace{0.15cm}
	\subfloat[Pruned AFT, annotated with event labels (ev1 -- ev10).]{\includegraphics[width=0.42\textwidth]{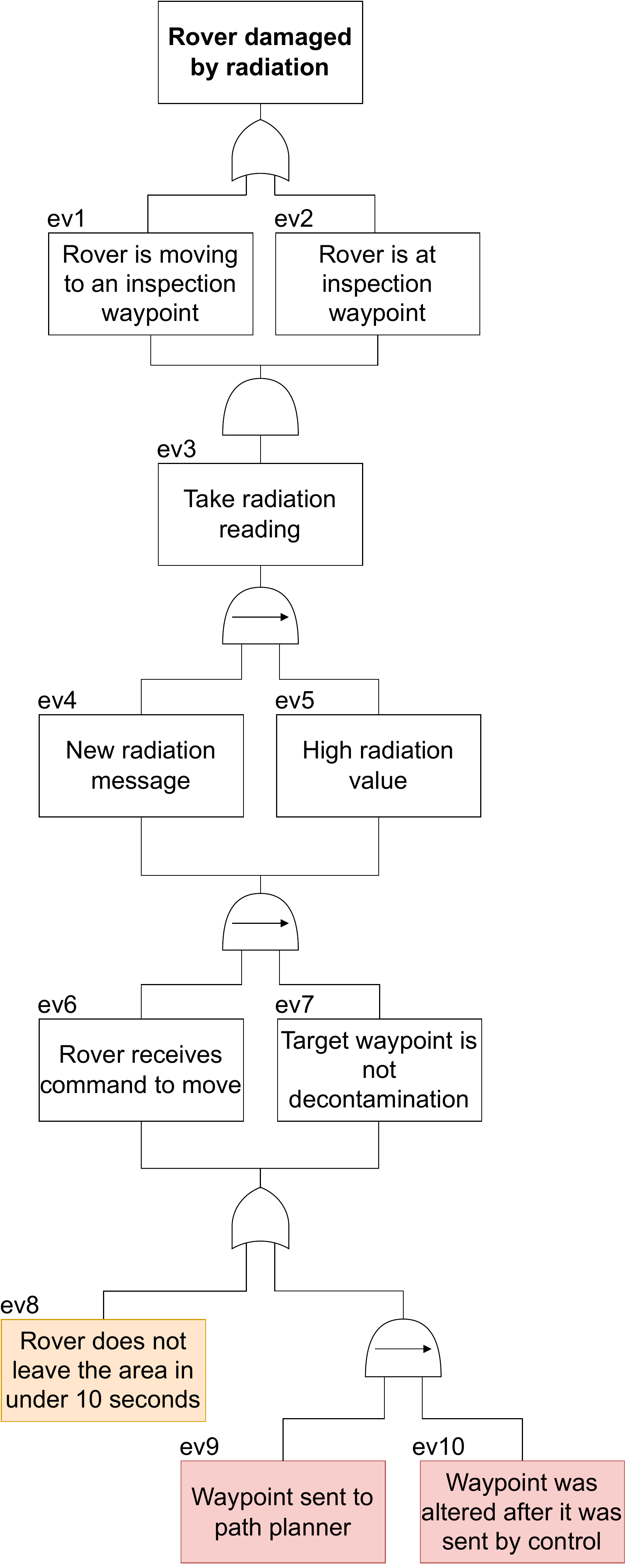}\label{fig:aftpruned}}
    \caption{An AFT for the remote inspection case study. Orange boxes represent faults, red boxes represent attacks.}
    \label{fig:aft-all}
\end{figure}

We show an example of an AFT for the remote inspection case study in Figure~\ref{fig:aft}. In this paper we focus on one particular undesired event, that a robot can be damaged by radiation. However, there could be more AFTs that specify security and safety issues for undesired events related to other parts of the system, such as dealing with collisions, battery, sensors, etc. The tree starts with a disjunction representing the two different possibilities of when the rover can be damaged by radiation. Either the rover is moving to an inspection waypoint, or it is already at the inspection waypoint. Both events eventually lead to the execution of the two actions that are available to the rover as part of this mission: taking radiation readings and collecting imagery. 
These actions are connected by an unordered $\mathit{AND}$ gate, meaning that there is no correct (or incorrect) order between them.

The ``take imagery'' event leads to a $\mathit{VOT}$ gate wherein $2$ out of the $3$ events are required to be true in order to trigger this branch of the tree. These events all describe possible faults when taking the images. Given that the environment is dynamic and noisy, we only consider the gate to be triggered if at least two of them occur. This branch of the tree does not cause direct radiation damage to the rover, but can be an indirect cause (e.g., the leakage is outside the range of the radiation sensor and should have been detected by taking images).

The other branches of the tree concern the events associated with taking a radiation reading using a radiation sensor. In this case, we have a $\mathit{SAND}$ gate to express that first a new radiation message must be observed and then the radiation value within that message must be high. This then leads to another $\mathit{SAND}$ gate where the rover receives a command to move to a target waypoint that is not the decontamination waypoint (i.e., the entrance). Finally, this leads to a disjunction with three possibilities: the rover is out of battery, the rover stays over 10 seconds in the contaminated area, or the contamination waypoint was sent to the planner but it was altered to be something else (by an ``attacker'').

We show an example of a pruned version of this AFT in Figure~\ref{fig:aftpruned}. We have pruned the branches about taking images since we assume they were verified using alternative verification techniques. We also remove the branch related to battery faults, since a battery model has not yet been implemented in the simulation. The pruned tree has four remaining branches (obtained from the two logical $\mathit{OR}$ gates), two result in a fault and the other two result in an attack.

\begin{table}[htbp]
\caption{Extending AFTs with runtime events and ROS topics.}\label{table:events}
\centering
\begin{tabular}{c|c|c}
\hline
AFT node & runtime event                       & ROS topic                            \\ \hline
ev1      & move(Waypoint)                      & /command                             \\
ev2      & movebase\_result(Waypoint, success) & /move\_base/result                   \\
ev3      & inspect(Waypoint)                   & /command                             \\
ev4      & radiation(Value, T1)                & /radiation\_sensor\_plugin/sensor\_0 \\
ev5      & Value $\geq$ 250           & -                                    \\
ev6      & move(NewWp, T2)                     & /command                             \\
ev7      & NewWp != entrance                   & -                                    \\
ev8      & T2 $\geq$ T1 + 10             & -                                    \\
ev9      & MoveBaseGoal(MBGoal)                & /move\_base/goal                     \\
ev10     & NewWp != MBGoal                     & -                                   
\end{tabular}
\end{table}

The runtime events associated with the RVAFT for the case study are shown in Table~\ref{table:events}. Besides adding the runtime event for each node, we also add the ROS topic where the event appears. This is only relevant because our case study is a ROS-based system and the tool we use to implement the monitors supports observing events by listening to specific ROS topics. Note that some nodes in the RVAFT do not have a topic listed (represented by -), this is because the parameters that are being tested are coming from previous nodes.


\subsection{Extracting Safety and Security Properties}
\label{sec:properties}

The following four properties are extracted from the four branches in the hierarchical decomposition of the RVAFT:\footnote{The $\land$ symbol is used to improve readability, but note that in RML the equivalent would be an empty space; in RML `$\land$'  denotes  intersection between two sets of events.}
$$
\begin{array}{l}
\varphi_1 = move(\mathit{Waypoint}) \;\land\; inspect(\mathit{Waypoint}) \;\land\; radiation(\mathit{Value}, T1) \;\land\; \\ \qquad \mathit{Value} \geq 250 \;\land\; move(\mathit{NewWP}, T2) \;\land\; \mathit{NewWP} \; != \; entrance \;\land\; T2 \geq T1 + 10  \\

\varphi_2 = movebase\_result(\mathit{Waypoint},success)  \;\land\; inspect(\mathit{Waypoint}) \;\land\; \\ \qquad radiation(\mathit{Value}, T1) \;\land\; \mathit{Value} \geq 250 \;\land\;  move(\mathit{NewWP}, T2) \;\land\; \\ \qquad \mathit{NewWP} \; != \; entrance \;\land\;  T2 \geq T1 + 10  \\

\varphi_3 = move(\mathit{Waypoint}) \;\land\; inspect(\mathit{Waypoint}) \;\land\; radiation(\mathit{Value}, T1) \;\land\; \\ \qquad \mathit{Value} \geq 250 \;\land\; move(\mathit{NewWP}, T2) \;\land\; \mathit{NewWP} \; != \; entrance \;\land\; \\ \qquad MoveBaseGoal(MBGoal) \;\land\; \mathit{NewWP} \; != \; MBGoal  \\

\varphi_4 = movebase\_result(\mathit{Waypoint},success)  \;\land\; inspect(\mathit{Waypoint}) \;\land\; \\ \qquad radiation(\mathit{Value}, T1) \;\land\; \mathit{Value} \geq 250 \;\land\;  move(\mathit{NewWP}, T2) \;\land\; \\ \qquad \mathit{NewWP} \; != \; entrance  \;\land\;  MoveBaseGoal(MBGoal) \;\land\; \mathit{NewWP} \; != \; MBGoal  \\
\end{array}
$$
with $\varphi_1$ and $\varphi_2$ representing the branches with the fault when the robot is moving to, or is at, an inspection waypoint, and $\varphi_3$ and $\varphi_4$ representing the branches with the attack when the robot is moving to, or is at, an inspection waypoint.
%
%
%
%
Since $\varphi_1$ differs from $\varphi_2$ only in the first observable event (the same for $\varphi_3$ and $\varphi_4$), we can merge them into one property and then go one step further and combine the resulting two properties into a single $\varphi$ property:
$$
\begin{array}{l}
\varphi = (move(\mathit{Waypoint}) \;\lor\; movebase\_result(\mathit{Waypoint},success)) \;\land\; \\ \qquad inspect(\mathit{Waypoint}) \;\land\; radiation(\mathit{Value}, T1) \;\land\; \mathit{Value} \geq 250 \;\land\; \\ \qquad  move(\mathit{NewWP}, T2) \;\land\; \mathit{NewWP} \; != \; entrance \;\land\; ((T2 \geq T1 + 10) \;\lor\; \\ \qquad (MoveBaseGoal(MBGoal) \;\land\; \mathit{NewWP} \; != \; MBGoal))
\end{array}
$$
where the disjunction symbols ($\lor$) represent the original branches in the tree.

Despite being able to generate such a property from the start, in our experience it was easier to generate individual properties for each branch and then merge them as desired. There are no practical benefits for doing one or the other, their differences are more qualitative such as readability and easiness to update the property if needed.
Note that while these properties were written with RML syntax, they still have to be converted into proper RML terms to be used by monitors.

\subsection{Generating the Monitors}
\label{sec:monitor}



To implement and run our monitors we use ROSMonitoring\footnote{\url{https://github.com/autonomy-and-verification-uol/ROSMonitoring} (Accessed: 08/09/2022)}~\cite{Ferrando20a}, a runtime verification framework developed to verify message passing in ROS applications. It supports the specification of properties in RML. Given formal properties to verify, ROSMonitoring generates monitors capable of perceiving messages exchanged amongst nodes in ROS. Such messages are then analysed and checked against the formal properties of interest. 

There are several options on how to generate monitors based on the properties described in Section~\ref{sec:properties}: (a) one monitor per RVAFT; (b) one monitor per branch; and (c) one monitor for faults and one monitor for attacks. 
Systems with multiple RVAFTs will benefit from option (a), since this can decrease the overhead from adding too many monitors. Options (b) and (c) are best used depending on how the recovery mechanism interacts with the monitor, for example, in option (c) we can have two different recovery mechanisms (faults and attacks) that require additional information from the monitors depending on their speciality. In this paper, we focus only on detection, thus, our example for monitor generation uses option (a).

To generate a monitor in ROSMonitoring we need a monitor configuration file and the properties specified in one of the supported formalisms (RML in our case). Generating the monitor configuration file is straightforward, we only need to know the topics where we can observe the events related to the property (which we can obtain from the RVAFT) and the nodes that publish in these topics (can be obtained from analysing the system).\footnote{Files available at: \url{https://github.com/rafaelcaue/rvaft-remote-inspection} (Accessed: 08/09/2022)} The RML specification for property $\varphi$ is shown in Listing~\ref{list:prop}. 
Note that in this complete RML specification we can reference which topic the event is related to, as well as the ROS message type of the parameters. 

\lstset{
	morekeywords={not,none,any,empty,matches,all,let,if,else,silent,id,topics,node,log,nodes,monitor,monitors,with},
	keywordstyle=\color{blue},
	morestring=[b]',
	stringstyle=\color{red},
	morecomment=[l]{//},
	commentstyle=\color{olive},
	mathescape=true,
	basicstyle=\scriptsize\ttfamily,
	captionpos=b,
	tabsize=4,
	breaklines,
	breakatwhitespace,
	showstringspaces=false,
	keepspaces
}

\begin{lstlisting} [float=t,caption={The RML specification for property $\varphi$ extracted from the RVAFT.},captionpos=b, label={list:prop}]
move(Waypoint) matches { topic: 'command', waypoint: Waypoint };
movebase_result(Waypoint, success) matches { topic: 'move_base/result', waypoint: Waypoint, result: success };
inspect(Waypoint) matches { topic: 'command', waypoint: Waypoint };
radiation(T1) matches  { topic: 'radiation_sensor_plugin/sensor_0', value: Value, time: T1 } with Value >= 250;
move(NewWp, T1) matches { topic: 'command', waypoint: NewWp, time: T2 } with NewWp != entrance and T2>=T1+10;
MoveBaseGoal(NewWp) matches { topic: 'move_base/goal', goal: MBGoal } with NewWp != MBGoal;
Main = { let Waypoint, NewWp, T1; (move(Waypoint) \/ movebase_result(Waypoint, success)) inspect(Waypoint) radiation(T1) (move(NewWp, T1) \/ (MoveBaseGoal(NewWp))) }
\end{lstlisting}


We evaluate our approach by running the monitor in the simulated environment from the case study. 
We show the event log of each monitor and the history of verdicts. Note that the properties extracted from the RVAFT represent `bad' scenarios (faults and attacks). This means that violations of these properties that are reported by the monitors are good outcomes, while satisfactions represent bad outcomes (a fault or attack has been detected). We also carry out some experiments to check the overhead of different configurations of monitors.

\subsection{Fault and Attack Detection}
\label{sec:eval}

As presented in Section~\ref{sec:properties}, given the tree of Figure~\ref{fig:aft-all}, we can extract the corresponding property $\varphi$. 
Using $\varphi$, we can automatically synthesise a runtime monitor by expressing $\varphi$ as an RML formula (reported in Listing~\ref{list:prop}). Such formula can be verified at runtime over finite traces of events generated through the system execution. As shown in Section~\ref{sec:properties}, $\varphi$ is the combination of four different properties $\varphi_i$ (with $1 \leq i \leq 4$). With each $\varphi_i$ derived by one of the four branches of the RVAFT. Since the four properties all share the same alphabet of events, their separation is not going to have a large influence on the kind of traces that they can handle; nonetheless, we report the results obtained by analysing execution traces by using four different monitors ($\varphi_1$ -- $\varphi_4$) as well as the combined property ($\varphi$).

\begin{table}[ht]
    \centering
        \caption{Example of `bad' traces (i.e., traces that satisfy the specifications).}
    \resizebox{\textwidth}{!}{
    \begin{tabular}{|c|c|c|c|c|c|}
    \hline
    \textbf{Events} & \textbf{$\varphi$} & \textbf{$\varphi_1$} & \textbf{$\varphi_2$} & \textbf{$\varphi_3$} & \textbf{$\varphi_4$} \\
    \hline
        \{``topic'': ``/command'', ``time'': 10.4, ``name'': ``move'', ``waypoint'': 0\} & $?$ & $?$ & & & \\
        \{``topic": ``/command", ``time": 15.6, ``name": ``inspect",``waypoint": 0\} & $?$ & $?$ & & & \\
        \{``pose": {$\ldots$}, ``value": 257.0, ``topic": ``/radiation\_sensor\_plugin/sensor\_0", ``time": 16.1 \} & $?$ & $?$ & & & \\
        \{``topic": ``/command", ``time": 30.241, ``name": ``move", ``waypoint": 1\} & $\top$ & $\top$ & & & \\
    \hline
    \textbf{Events} & \textbf{$\varphi$} & \textbf{$\varphi_1$} & \textbf{$\varphi_2$} & \textbf{$\varphi_3$} & \textbf{$\varphi_4$} \\
    \hline
       \{``topic'': ``/move\_base/result'', ``time'': 8.2, ``waypoint'': 0, ``result": 'success'\} & $?$ & & $?$ & &  \\
        \{``topic": ``/command", ``time": 15.6, ``name": ``inspect",``waypoint": 0\} & $?$ & & $?$ & &  \\
        \{``pose": {$\ldots$}, ``value": 257.0, ``topic": ``/radiation\_sensor\_plugin/sensor\_0", ``time": 16.1 \} & $?$ & & $?$ & &  \\
        \{``topic": ``/command", ``time": 30.493, ``name": ``move", ``waypoint": 1\} & $\top$ & & $\top$ & &  \\
    \hline
    \textbf{Events} & \textbf{$\varphi$} & \textbf{$\varphi_1$} & \textbf{$\varphi_2$} & \textbf{$\varphi_3$} & \textbf{$\varphi_4$} \\
    \hline
        \{``topic'': ``/command'', ``time'': 8.2, ``name'': ``move'', ``waypoint'': 0\} & $?$ & & & $?$ &  \\
        \{``topic": ``/command", ``time": 12.6, ``name": ``inspect",``waypoint": 0\} & $?$ & & & $?$ &  \\
        \{``pose": {$\ldots$}, ``value": 257.0, ``topic": ``/radiation\_sensor\_plugin/sensor\_0", ``time": 14.1 \} & $?$ & & & $?$ &  \\
        \{"topic": ``/move\_base/goal", ``goal": 2, ``time": 22.405\} & $\top$ & & & $\top$ &  \\
    \hline
    \textbf{Events} & \textbf{$\varphi$} & \textbf{$\varphi_1$} & \textbf{$\varphi_2$} & \textbf{$\varphi_3$} & \textbf{$\varphi_4$} \\
    \hline
        \{``topic'': ``/move\_base/result'', ``time'': 8.2, ``waypoint'': 0, ``result": 'success'\} & $?$ & & & & $?$ \\
        \{``topic": ``/command", ``time": 12.6, ``name": ``inspect",``waypoint": 0\} & $?$ & & & & $?$ \\
        \{``pose": {$\ldots$}, ``value": 257.0, ``topic": ``/radiation\_sensor\_plugin/sensor\_0", ``time": 14.1 \} & $?$ & & & & $?$ \\
        \{"topic": ``/move\_base/goal", ``goal": 2, ``time": 22.405\} & $\top$ & & & & $\top$ \\
    \hline
    \end{tabular}
    }

    \label{tab:phi-sat}
\end{table}

\begin{table}[ht]
    \centering
    \caption{Example of `good' traces (i.e., traces that violate the specifications).}
    \resizebox{\textwidth}{!}{
            
    \begin{tabular}{|c|c|c|c|c|c|}
    \hline
    \textbf{Events} & \textbf{$\varphi$} & \textbf{$\varphi_1$} & \textbf{$\varphi_2$} & \textbf{$\varphi_3$} & \textbf{$\varphi_4$} \\
    \hline
        \{``topic'': ``/command'', ``time'': 10.4, ``name'': ``move'', ``waypoint'': 0\} & $?$ & $?$ & & & \\
        \{``topic": ``/command", ``time": 15.6, ``name": ``inspect",``waypoint": 0\} & $?$ & $?$ & & & \\
        \{``pose": {$\ldots$}, ``value": 257.0, ``topic": ``/radiation\_sensor\_plugin/sensor\_0", ``time": 16.1 \} & $?$ & $?$ & & & \\
        \{``topic": ``/command", ``time": 17.493, ``name": ``move", ``waypoint": 1\} & $\bot$ & $\bot$ & & & \\
    \hline
    \textbf{Events} & \textbf{$\varphi$} & \textbf{$\varphi_1$} & \textbf{$\varphi_2$} & \textbf{$\varphi_3$} & \textbf{$\varphi_4$} \\
    \hline
      \{``topic'': ``/move\_base/result'', ``time'': 8.2, ``waypoint'': 0, ``result": 'success'\} & $?$ & & $?$ & & \\
        \{``topic": ``/command", ``time": 15.6, ``name": ``inspect",``waypoint": 0\} & $?$ & & $?$ & & \\
        \{``pose": {$\ldots$}, ``value": 257.0, ``topic": ``/radiation\_sensor\_plugin/sensor\_0", ``time": 16.1 \} & $?$ & & $?$ & & \\
        \{``topic": ``/command", ``time": 17.493, ``name": ``move", ``waypoint": 1\} & $\bot$ & & $\bot$ & & \\
    \hline
    \textbf{Events} & \textbf{$\varphi$} & \textbf{$\varphi_1$} & \textbf{$\varphi_2$} & \textbf{$\varphi_3$} & \textbf{$\varphi_4$} \\
    \hline
        \{``topic'': ``/command'', ``time'': 8.2, ``name'': ``move'', ``waypoint'': 0\} & $?$ & & & $?$ & \\
        \{``topic": ``/command", ``time": 12.6, ``name": ``inspect",``waypoint": 0\} & $?$ & & & $?$ & \\
        \{``pose": {$\ldots$}, ``value": 257.0, ``topic": ``/radiation\_sensor\_plugin/sensor\_0", ``time": 14.1 \} & $?$ & & & $?$ & \\
        \{"topic": ``/move\_base/goal", ``goal": 1, ``time": 22.405\} & $\bot$ & & & $\bot$ & \\
    \hline
    \textbf{Events} & \textbf{$\varphi$} & \textbf{$\varphi_1$} & \textbf{$\varphi_2$} & \textbf{$\varphi_3$} & \textbf{$\varphi_4$} \\
    \hline
        \{``topic'': ``/move\_base/result'', ``time'': 8.2, ``waypoint'': 0, ``result": 'success'\} & $?$ & & & & $?$ \\
        \{``topic": ``/command", ``time": 12.6, ``name": ``inspect",``waypoint": 0\} & $?$ & & & & $?$ \\
        \{``pose": {$\ldots$}, ``value": 257.0, ``topic": ``/radiation\_sensor\_plugin/sensor\_0", ``time": 14.1 \} & $?$ & & & & $?$ \\
        \{"topic": ``/move\_base/goal", ``goal": 1, ``time": 22.405\} & $\bot$ & & & & $\bot$ \\
    \hline
    \end{tabular}
    }

    \label{tab:phi-vio}
\end{table}

In Table~\ref{tab:phi-sat} we report the results obtained by analysing some finite traces collected while running the system under analysis. In particular, we report traces of events that lead to the satisfaction of the property, i.e., that represent when an attack or a fault have been identified. In the first column, we have four different sets of traces (one example per $\varphi_i$, with empty cells when the trace is not covered by the corresponding property) that are analysed and the verification process concludes the satisfaction of $\varphi$. Note that, the resulting runtime monitor reports $?$ as long as it is uncertain on the satisfaction/violation ($\top$ and $\bot$, respectively) of $\varphi$. 

Each event contains information about the system. For instance, the first event in the first trace corresponds to observing the system performing a command ``move'', towards waypoint 0, at time $10.4sec$ (w.r.t. when the mission started). In this example trace, the radiation level perceived is 257, which is greater than 250 (and so is assumed dangerous for the robot); hence, the robot is supposed to leave the waypoint in less than $10sec$ to reduce the chance of being damaged. Since the next command to move to another waypoint is performed at time $30.241sec$, which is greater than $10sec$ since when the radiation was perceived ($30.241sec - 16.1sec = 14.141sec$), the property is being satisfied which means that the monitor has identified a potential fault. This particular property represents the case where the rover is moving to an inspection waypoint and has taken a high radiation reading, but it has not moved away from the area on time. The same reasoning applies for the other traces. In Table~\ref{tab:phi-vio}, we show examples of traces that violate the properties, i.e., traces where no attack or violation have been detected.

The RVAFT adds an additional layer of explainability that helps the user to interpret the potential causes that led to a failure or an attack. Instead of having to parse the events that led to the satisfaction/violation of the property, the user can consult the respective branch in the RVAFT, which is easy to do since each node in the RVAFT is assigned to a particular runtime event. Further exploration of explainability is out of scope for this paper, but it may prove to be useful in future work when considering runtime enforcement.

\subsection{Runtime Monitors Overhead}

When working with RV, it is important to evaluate the possible overhead introduced by the monitors at runtime. Even though runtime monitors are known to be lightweight, they can still cause some computational impact. Since the system may have a limited amount of resources, to estimate the overhead of such verification process is crucial. We carried out several experiments running the remote inspection simulation with, and without, runtime monitors associated to the case study AFT, and we obtained almost no overhead. The specifications of the hardware used for these experiments were as follows: Intel(R) Core(TM) i7-7700HQ CPU @ 2.80GHz, 4 cores 8 threads, 16 GB RAM DDR4.

\begin{figure}[ht]
    \centering
    \includegraphics[width=0.7\linewidth]{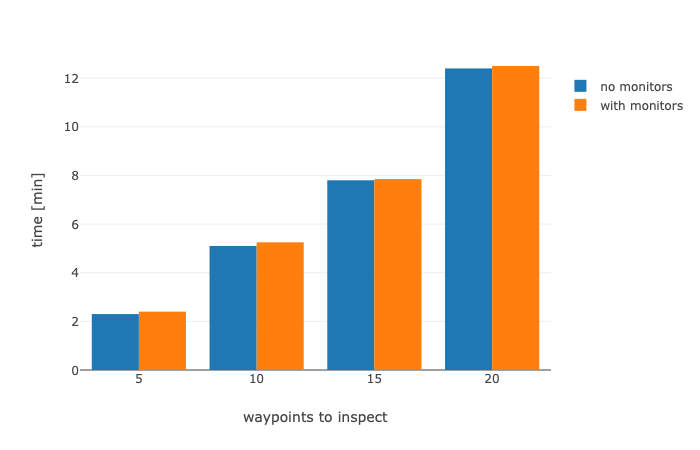}
    \caption{Overhead obtained by adding runtime monitors derived from AFT to the case study.}
    \label{fig:overhead}
\end{figure}

In Figure~\ref{fig:overhead}, we report the results for varying the number of waypoints for the rover to inspect. Increasing the number of waypoints to inspect will increase the time required by the rover to complete the mission, however, the overhead introduced by the monitors remains almost constant. 
Because the monitors are external to the system and do not interfere with it (for instance, no runtime enforcement is attempted), we expect similar negligible overhead for larger and more complex scenarios as well. More details on the performance cost and overhead of monitors in ROSMonitoring can be found in~\cite{Ferrando20a}.


\section{Conclusion}
\label{sec:conc}

In this paper, we have introduced RVAFTs, an extension of AFTs with runtime events. RVAFTs can be used to generate monitors that are capable of performing runtime verification to detect when the branches from the RVAFT occur at runtime. In other words, monitors are generated to check for all possible faults and attacks that are described in the RVAFTs. 
We have demonstrated the use of our approach in a robotic system, how an initial AFT can be extended to a RVAFT, how runtime monitors can be generated from it, and we have also shown how this can be achieved using a practical tool (ROSMonitoring) by translating RVAFTs into RML specifications. Detection is the first part in trying to avoid undesirable events from an RVAFT, future work is also necessary to trigger the appropriate recovery mechanism. 

As future work we plan to extend our monitors to be able to trigger the relevant recovery mechanism depending on which branch was detected, attack or fault, as well as expanding the information the monitor reports based on the RVAFT. Another avenue for further work is to make use of the pre-deployment statistical model checking techniques for AFTs in~\cite{aft} in order help prune branches that are considered less likely or less dangerous in the AFT before extending it to a RVAFT. 
Additional gates could also be introduced to allow RVAFTs to deal with temporal events (simultaneous or interval based events). And finally, a future extension of this work is to generate translations of RVAFTs to multi-path logics, such as HyperLTL~\cite{DBLP:conf/post/ClarksonFKMRS14}.


\bibliographystyle{eptcs}
\bibliography{main}

\end{document}